\documentclass[a4paper,12pt]{elsarticle}

\usepackage{lineno,hyperref} %colordvi}
\usepackage{color}                        % For color text: \color
\usepackage{xspace}
\usepackage{pdfwidgets}
\usepackage{amssymb}                    % useful mathematical symbols
\usepackage{pdflscape}

\modulolinenumbers[5]

\journal{JASTP}

\biboptions{authoryear}

\newcommand{\ie}{{\it i.e.~}}
\newcommand{\eg}{{\it e.g.}}

\chardef\us=`\_

\begin{document}

\begin{frontmatter}

\title{Magnetic structure of solar flare regions producing hard X-ray pulsations}
%\title{Elsevier \LaTeX\ template\tnoteref{mytitlenote}}





\author[aff1,aff2,aff3]{I.V.~Zimovets\corref{mycorrespondingauthor}}
\cortext[mycorrespondingauthor]{Corresponding author}
\ead{ivanzim@iki.rssi.ru}

\author[aff1]{R.~Wang}
\author[aff1,aff4]{Y.D.~Liu}
\author[aff1]{C.~Wang}
\author[aff5]{S.A.~Kuznetsov}
\author[aff3,aff6]{I.N.~Sharykin}
\author[aff3,aff7]{A.B.~Struminsky}
\author[aff8,aff9]{V.M.~Nakariakov}

\address[aff1]{State Key Laboratory of Space Weather, National Space Science Center (NSSC) of the Chinese Academy of Sciences, No.1 Nanertiao, Zhongguancun, Haidian District, Beijing 100190, China  }
\address[aff2]{International Space Science Institute -- Beijing (ISSI-BJ), No.1 Nanertiao, Zhongguancun, Haidian District, Beijing 100190, China }
\address[aff3]{Space Research Institute (IKI) of the Russian Academy of Sciences, Profsoyuznaya str. 84/32, Moscow 117997, Russia   }
\address[aff4]{University of Chinese Academy of Sciences, Beijing 196140, China }
\address[aff5]{Central Astronomical Observatory at Pulkovo of the Russian Academy of Sciences, Pulkovskoye chaussee 65/1, Saint-Petersburg 196140, Russia }
\address[aff6]{Institute of Solar-Terrestrial Physics, Russian Academy of Sciences, Siberian Branch, Lermontov st., 126a, Irkutsk p/o box 291 664033, Russia }
\address[aff7]{Moscow Institute of Physics and Technology (State University), Institutskyi per. 9, Dolgoprudny, Moscow Region 141700, Russia   }
\address[aff8]{Centre for Fusion, Space and Astrophysics, Department of Physics, University of Warwick, Coventry CV4 7AL, UK}
\address[aff9]{St. Petersburg Branch, Special Astrophysical Observatory, Russian Academy of Sciences, St. Petersburg 196140, Russia}

\begin{abstract}
We present analysis of the magnetic field in seven solar flare regions accompanied by the pulsations of hard X-ray (HXR) emission. These flares were studied by \cite{2016SoPh..291.3385K} (Paper~I), and chosen here because of the availability of the vector magnetograms for their parent active regions (ARs) obtained with the SDO/HMI data. In Paper~I, based on the observations only, it was suggested that a magnetic flux rope (MFR) might play an important role in the process of generation of the HXR pulsations. The goal of the present paper is to test this hypothesis by using the extrapolation of magnetic field with the non-linear force-free field (NLFFF) method. Having done this, we found that before each flare indeed there was an MFR elongated along and above a magnetic polarity inversion line (MPIL) on the photosphere. In two flare regions the sources of the HXR pulsations were located at the footpoints of different magnetic field lines wrapping around the central axis, and constituting an MFR by themselves. In five other flares the parent field lines of the HXR pulsations were not a part of an MFR, but surrounded it in the form of an arcade of magnetic loops. These results show that, at least in the analyzed cases, the ``single flare loop'' models do not satisfy the observations and magnetic field modeling, while are consistent with the concept that the HXR pulsations are a consequence of successive episodes of energy release and electron acceleration in different magnetic flux tubes (loops) of a complex AR. An MFR could generate HXR pulsations by triggering episodes of magnetic reconnection in different loops in the course of its non-uniform evolution along an MPIL. However, since three events studied here were confined flares, actual eruptions may not be required to trigger sequential particle acceleration episodes in the magnetic systems containing an MFR.
\end{abstract}
%%%%%


\end{frontmatter}

%\linenumbers

%%%%%%%%%%%%%%%%%%%%%%%%%%%%%%%%%%%%%%%%%%%%%%%%%%%
\section{Introduction}\label{S-Intro} 

The processes of energy release in solar flares, especially in the impulsive phase, usually are intermittent and non-stationary. This is well evidenced by the presence of multiple peaks (bursts or pulsations) of different amplitudes and duration in the light curves of flare electromagnetic radiation in a broad range of wavelengths: from radio waves to hard X-rays (HXRs), and sometimes even up to gamma-rays (\citealp{1988SoPh..118...49D,2002SSRv..101....1A,2007ApJ...662..691M,2010SoPh..267..329K,2010ApJ...708L..47N,2015SoPh..290.3625S}). Stellar flares often show similar properties (\eg, see the discussion of ``complex'' flares in \cite{2014ApJ...797..122D}).

Despite many years of studying of flare quasi-periodic pulsations (QPP), there is no full understanding of the underlying physical mechanisms yet (\citealp{2010PPCF...52l4009N,2016SSRv..200...75N,2016SoPh..291.3143V,2018SSRv..214...45M}). In general, it is believed that the energy in solar flares is released by means of magnetic reconnection (\citealp{2002A&ARv..10..313P,2011LRSP....8....6S}). Most probably, flare pulsations are also associated someway to magnetic reconnection. There are two main groups of possible models of long-period pulsations ($P \gtrsim 1$~s), which are the main subject of the present work: (1) based on MHD waves and oscillations, including the wave-driven reconnection; (2) based on the so-called ``load/unload'' mechanisms, \ie spontaneous repetitive magnetic reconnection (\citealp{2009SSRv..149..119N}). 

The first group of models is more popular because of the direct link of the observed quasi-periodicity with the periodicity of the wave processes, ubiquity of MHD waves and oscillations in the solar atmosphere, and their potential ability to influence all main aspects of the generation of electromagnetic emission in flare regions. In particular, MHD oscillations and waves can be a quasi-periodic trigger/modulator of magnetic reconnection and can influence dynamics of non-thermal particles and plasma in flare loops. Moreover, the MHD oscillations based models are attractive since they could help to diagnose physical parameters of the flaring site (such as plasma density and magnetic field), if there is confidence in the correct choice of the model used (\eg, \citealp{2014SoPh..289.3233L,2016SSRv..200...75N}). 

The ``load/unload'' mechanisms are mainly based on the possibility of the repetitive regimes of energy release in the flare sites through the ``bursty'' magnetic reconnection (\citealp{2000A&A...360..715K}), associated with successive generation of multiple magnetic islands and their subsequent coalescence in an extended quasi-vertical macroscopic current sheet generated in course of a flare development (\citealp{2001EP&S...53..473S,2006Natur.443..553D,2010A&A...514A..28K}). There are also several other models belonging to this group and based on different ideas (see, \cite{2009SSRv..149..119N,2016SSRv..200...75N,2016SoPh..291.3143V}, as reviews on this issue). 

Possibly, different mechanisms can operate in different flares, due to the wide variety of the physical processes included in the flare physics, or different mechanisms can accompany one another in the same flare region. Spatially-resolved observations of sources of the flare pulsations are important for understanding their mechanisms, and for reliable identification of the models used for their interpretation (\eg, \citealp{2005ApJ...625L.143G,2009SoPh..258...69Z,2012ApJ...748..139I,2013AstL...39..267Z,2014SoPh..289.1239N,2015ApJ...804L...8L,2017ApJ...836...84D}). 

Recently, based on the systematic analysis of spatially-resolved observations made by RHESSI (\citealp{2002SoPh..210....3L}) it was shown that footpoint (chromospheric) sources of HXR pulsations (with time differences between successive HXR peaks within the range $P \approx 8-270$~s) in all (29) flares studied are not stationary (anchored) in space~--- they demonstrate apparent displacement in the parent active regions (ARs) from pulsation to pulsation (\citealp{2016SoPh..291.3385K}, hereafter referred to as Paper~I). Based on these observations, it was concluded that the mechanism of flare HXR pulsations (at least with the characteristic time differences between the successive peaks $P$ in the considered range) is related to successive triggering of the flare energy release in different magnetic loops of the parent ARs. The triggering mechanism was not directly identified in Paper~I. Based on the fact that more than $85\%$ of the analyzed flares were accompanied by coronal mass ejections (CMEs), \ie were eruptive events, it was assumed that a non-uniformly erupting magnetic flux rope (MFR) could act as a trigger of the flare energy release. Successive interaction of different parts of the MFR with certain, spatially separated loops of the parent active region could initiate episodes of spatially-localized magnetic reconnection and acceleration of electrons, and, as a result, could lead to apparent motion of the HXR sources and to a series of the HXR pulsations. However, in Paper~I the presence of an MFR in the parent ARs before the flares was just hypothesized, but it was not confirmed either by observations, or by extrapolation of the magnetic field. 

The goal of the present paper is to investigate the geometry (structure) of the magnetic field in the flare regions studied in Paper~I, based on the reconstruction (extrapolation) of the magnetic field in the non-linear force-free field (NLFFF) approximation (\citealp{2012LRSP....9....5W}). The first task is to verify whether MFRs were indeed presented in those ARs prior to the flare onset or not. It is known that an MFR can be present in an AR before its eruption and the subsequent flare, or an MFR can be formed from a sheared arcade due to the magnetic reconnection (\citealp{2002A&ARv..10..313P,2013AdSpR..51.1967S,2017ScChE..60.1383C,2017ScChE..60.1408G}). We aim to check which of these two possibilities were realized in the ARs studied. To the best of our knowledge, such an analysis has not been performed systematically for flares with HXR pulsations. The second task is to analyze the spatial relation of MFRs (if present) and the parent magnetic field lines of the sources of the HXR pulsations. This will help to corroborate the aforementioned hypothesis on the important role of MFRs in generation of the flare HXR pulsations. Also, this will help to demonstrate explicitly that different HXR pulsations are emitted from different parts of an MFR rather than from a ``single'' oscillating loop as it is often assumed in some models of flare pulsations (\eg, \citealp{2008PhyU...51.1123Z,2010SoPh..267..329K,2016ApJ...833..114C}).  

The paper is organized as follows. Selection and analysis (magnetic field extrapolation, visualization of the sources of HXR pulsations and magnetic field lines) of the flare regions is described in Section~\ref{s-da}. The main results of the analysis are summarized and discussed in Section~\ref{S-sumres}. Conclusions are given in Section~\ref{S-Concl}.
%%%%%%%%%%%%%%%%%%%%%%%%%%%%%%%%%%%%%%%%%%%%%%%%%%%

%%%%%%%%%%%%%%%%%%%%%%%%%%%%%%%%%%%%%%%%%%%%%%%%%%%
\section{Data analysis}\label{s-da} 

%%%%%%%%%%%%%%%%%%%%%%%%%%%%%%%%%%%%%%%%%%%%%%%%%%%
\subsection{Selection of events}\label{ss-es}

\begin{figure} 
\centerline{\includegraphics[width=0.95\textwidth]{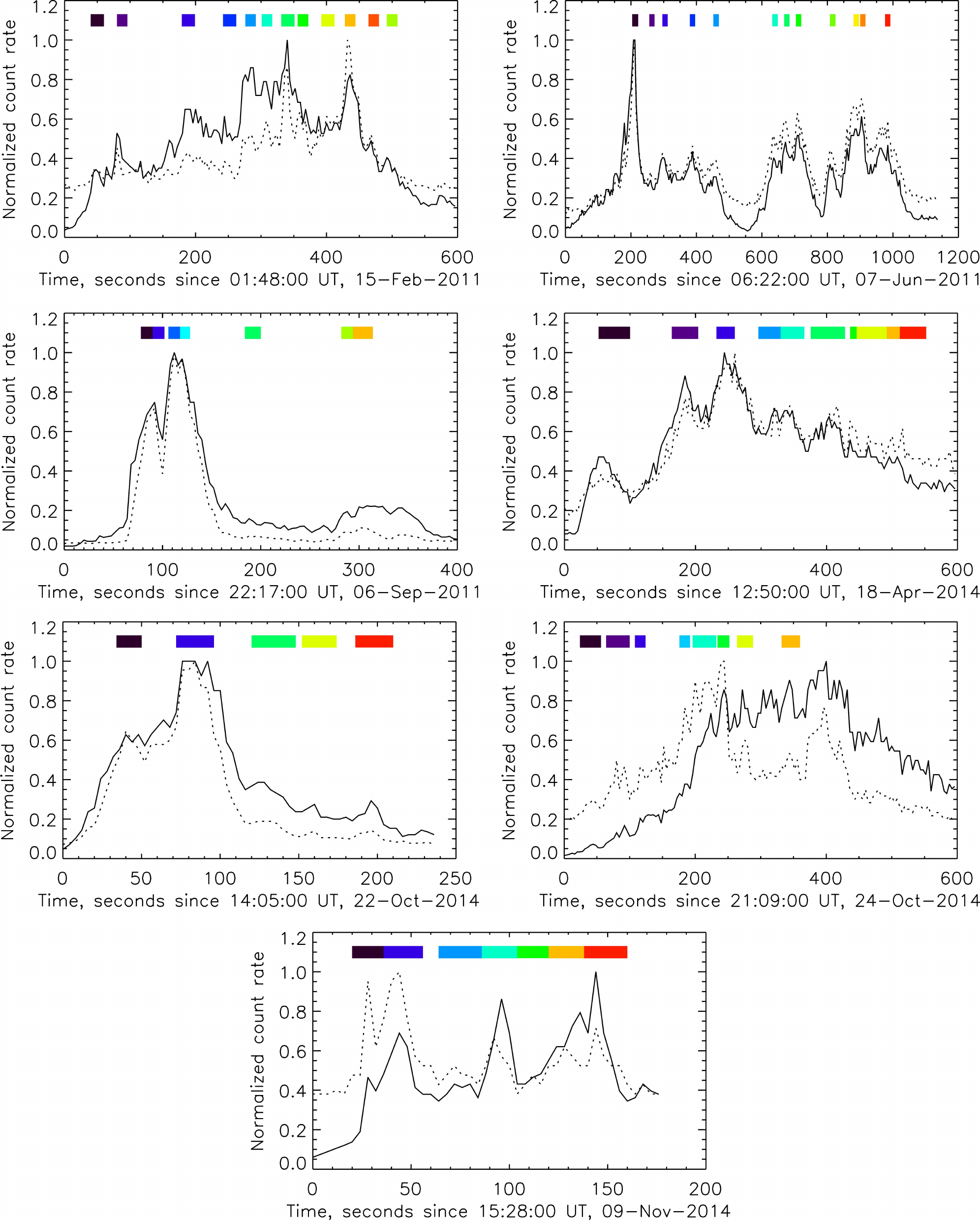}}

\caption{Normalized (to the maximum) four-second RHESSI corrected count rates in the $25-50$~keV (solid curves) and $50-100$~keV (dotted curves) energy channels, for seven solar flares studied. The color horizontal segments at the top mark the time intervals of different HXR pulsations for which the HXR images were synthesized and positions of the HXR sources were found for the further analysis. These time intervals (and corresponding colors) almost coincide with the time intervals (and corresponding colors) determined in Paper~I. Spatial positions of the centers of maximum brightness of these HXR sources are used as the starting points in the chromosphere for the reconstruction of the magnetic field lines shown in Figures~\ref{fig:fl1}--\ref{fig:fl2} by appropriate colors. These maximum brightness centers of the HXR sources are also shown in Figure~\ref{fig:HXRonly}.} 
\label{fig:HXRlc}
\end{figure}

For the analysis we took the last seven solar flares from the Paper~I catalog (No~23--29): SOL2011-02-15, SOL2011-06-07, SOL2011-09-06, SOL2014-04-18, SOL2014-10-22, SOL2014-10-24, and SOL2014-11-09 (see Table~\ref{tab:nlfff}). This choice is determined by the fact that the parent ARs of these events were observed by the Helioseismic and Magnetic Imager (HMI; \citealp{2012SoPh..275..207S,2012SoPh..275..229S}) instrument onboard the Solar Dynamics Observatory (SDO), and the vector magnetograms are available for these ARs, which is crucial for our study. We emphasize that this choice is determined only by the availability of these data, and by the events selection criteria in Paper~I. No other additional (subjective) criteria are used. The light curves of solar HXR emission detected by RHESSI during these flares are shown in Figure~\ref{fig:HXRlc}.
%%%%%%%%%%%%%%%%%%%%%%%%%%%%%%%%%%%%%%%%%%%%%%%%%%%

%%%%%%%%%%%%%%%%%%%%%%%%%%%%%%%%%%%%%%%%%%%%%%%%%%%
\subsection{Extrapolation of magnetic field}\label{ss-mfr}

For each of seven selected ARs we determined the pre-flare coronal magnetic field topology by adopting the non-linear force-free field (NLFFF) method developed by \citep{2000ApJ...540.1150W}, and extended by \citep{2004SoPh..219...87W} and \citep{2010A&A...516A.107W}. A pre-processing procedure \citep{2006SoPh..233..215W} was employed to remove most of the net force and torque from the data, so that the boundary can be more consistent with the force-free assumption. The NLFFF extrapolation used in our works adopts the same free parameters as Case-E in (\citealp{2012SoPh..281...37W}). As the boundary conditions, we used the Space-weather HMI Active Region Patches (SHARPs) data product described by \cite{2014SoPh..289.3549B}. This data has a time step of 12 minutes, similar to the standard full-disk SDO/HMI vector magnetograms. We chose the data for the instants of time before the flares, within $2-31$~minutes prior to the flare onset times according to the GOES data (see Table~1 in Paper-I). It is known that the magnetic field reconstructed in the NLFF approximation usually does not change significantly on a time scale of a few tens of minutes (\eg, \citealp{2008ApJ...679.1629G,2016ApJ...818..148L,2017ScChE..60.1408G}). The selected regions have a rectangular shape in the helioprojective-cartesian (HPC) coordinates (\citealp{2006A&A...449..791T}). Their angular sizes range from 160 to 300 arcseconds along the X axis, and from 128 to 256 arcseconds along the Y axis, to include all the main sources of the magnetic field in the flare area studied. For the extrapolation in six out of all seven events, we binned the data to $\approx 1.0^{\prime \prime}$ per pixel, except for the SOL2014-11-09 event, for which we kept the original pixel size of the HMI magnetograms of $\approx 0.5^{\prime \prime}$. This was done to decrease computational time, though without significant loss of quality. 

The calculations were performed in rectangular Cartesian coordinates, neglecting the sphericity of the photosphere. The inaccuracies associated with this approach are acceptable (see \citealp{1990SoPh..126...21G}), since the regions of interest selected for the magnetic field reconstruction were chosen small enough. Moreover, we are only interested in the central sections of the selected regions, where the studied flares occurred. The magnetic field near the edges, where larger errors are accumulated, is not of interest for this study. Almost all analyzed ARs are located near the center of the solar disk (see Table~1 in Paper~I). Only for the SOL2011-06-07 flare that happened at the helio-longitude of $\approx 50^{\circ}$, the expected inaccuracies are quite high ($\approx 4-6$ grid points) and should be taken into account.

In Table~\ref{tab:nlfff} we show the parameters used in the estimation of the quality of the non-linear force-free field reconstruction for the investigated flare regions. The parameters $L_{1}$ and $L_{2}$ are the measures of force- and divergence-freeness, respectively; the $\left(\textbf{j},\textbf{B}\right)$-angle is the angle between electric current density and magnetic field vectors averaged over a studied region; and $E_{nlff}/E_{pot}$ is the ratio of non-linear force-free and potential field energies in a studied region (\citealp{2012SoPh..281...37W}). The values of $L_{1}$ and $L_{2}$ are less than 2 for all the flares studied, indicating reasonable (relying on the experience of the previous tests) force- and divergence-freeness of the reconstructed fields. This is also confirmed by the low values ($<10^{\circ}$) of the $\left(\textbf{j},\textbf{B}\right)$-angle. The values of $E_{nlff}/E_{pot}$ are larger than 1 for all events that also indicates the adequacy of the magnetic reconstruction done.
%%%%%%%%%%%%%%%%%%%%%%%%%%%%%%%%%%%%%%%%%%%%%%%%%%%

%%%%%%%%%%%%%%%%%%%%%%%%%%%%%%%%%%%%%%%%%%%%%%%%%%%
\subsection{Visualization of the HXR sources and magnetic field lines}\label{ss-sup} 

\begin{figure} 
\centerline{\includegraphics[width=0.93\textwidth]{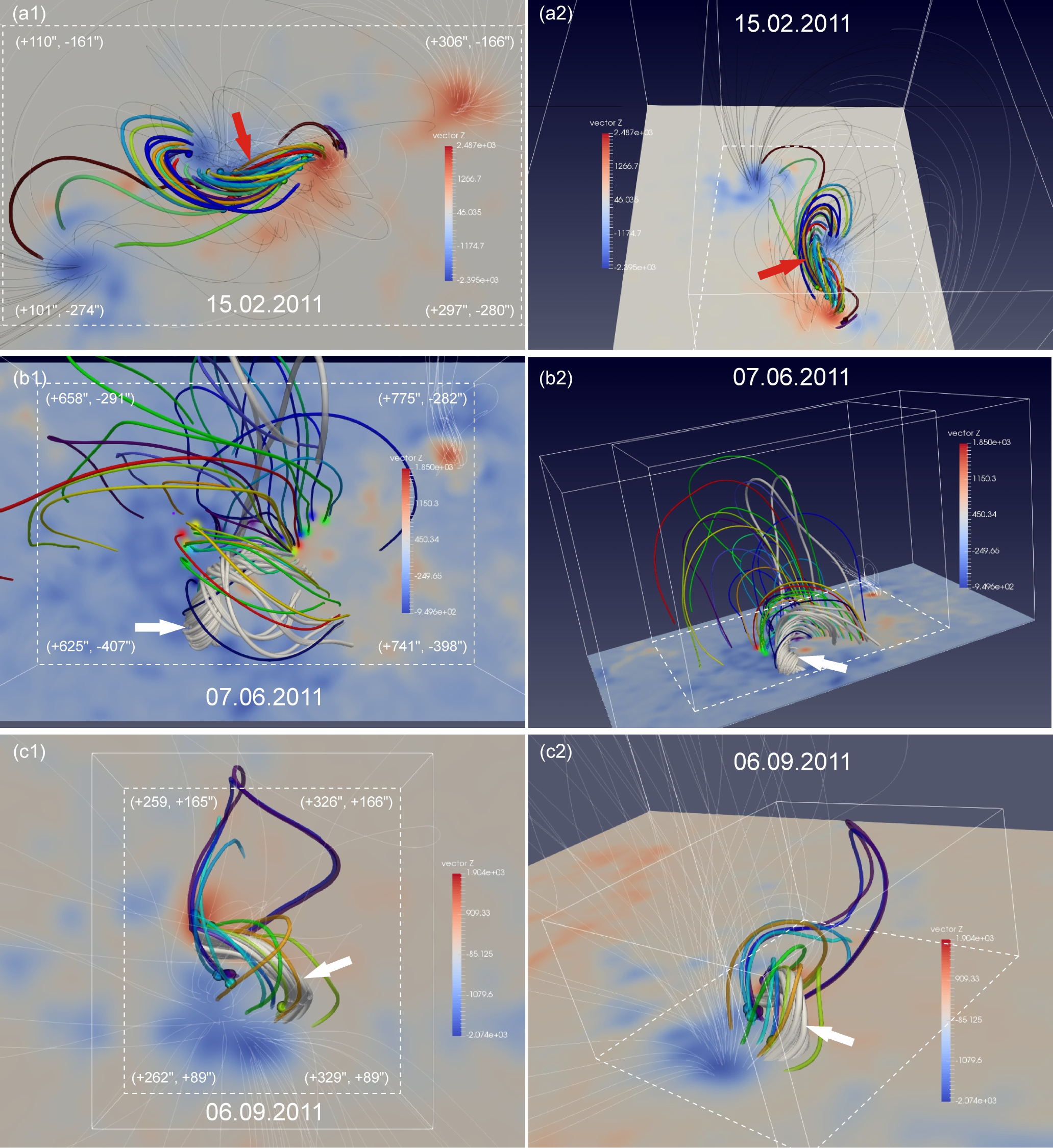}}

\caption{Reconstructed magnetic field lines in the regions of the 15-Feb-2011 (a), 07-Jun-2011 (b) and 06-Sep-2011 (c) solar flares accompanied by the HXR (25--50~keV) pulsations (see Figure~\ref{fig:HXRlc}). The top view is on the left, the side view is on the right. Colors of the field lines correspond to the colors of the sources of the HXR pulsations (see Figure~\ref{fig:HXRlc}; the spatial locations of these HXR sources are also shown in Paper~I). The HXR sources (\ie the positions of their brightness maxima) are shown here by small circles of the appropriate colors (see also Figure~\ref{fig:HXRonly}). The twisted bundles of the thick light gray field lines represent a magnetic flux rope (MFR) found in the core of these flare regions. The MFR is also indicated by the thick white arrow. The red arrow indicates the MFR composed of the field lines (colored) reconstructed from the sources of the HXR pulsations. Thin gray field lines are background magnetic field lines started from the strongest nearby magnetic sources. The background images are the maps of the radial  magnetic field component ($B_{r}$) on the photosphere made with the pre-flare SDO/HMI magnetograms (SHARPs). The colorbars show values of $B_{r}$ for the corresponding colors. The thin white dashed quadrangles left and right on the photosphere are shown just to indicate the regions of interest and to give the reference helioprojective cartesian coordinates (HPC) of its corners (in arcseconds). The same regions are also shown in Figure~\ref{fig:HXRlc}.} 
\label{fig:fl1}
\end{figure}

\begin{figure} 
\centerline{\includegraphics[width=0.95\textwidth]{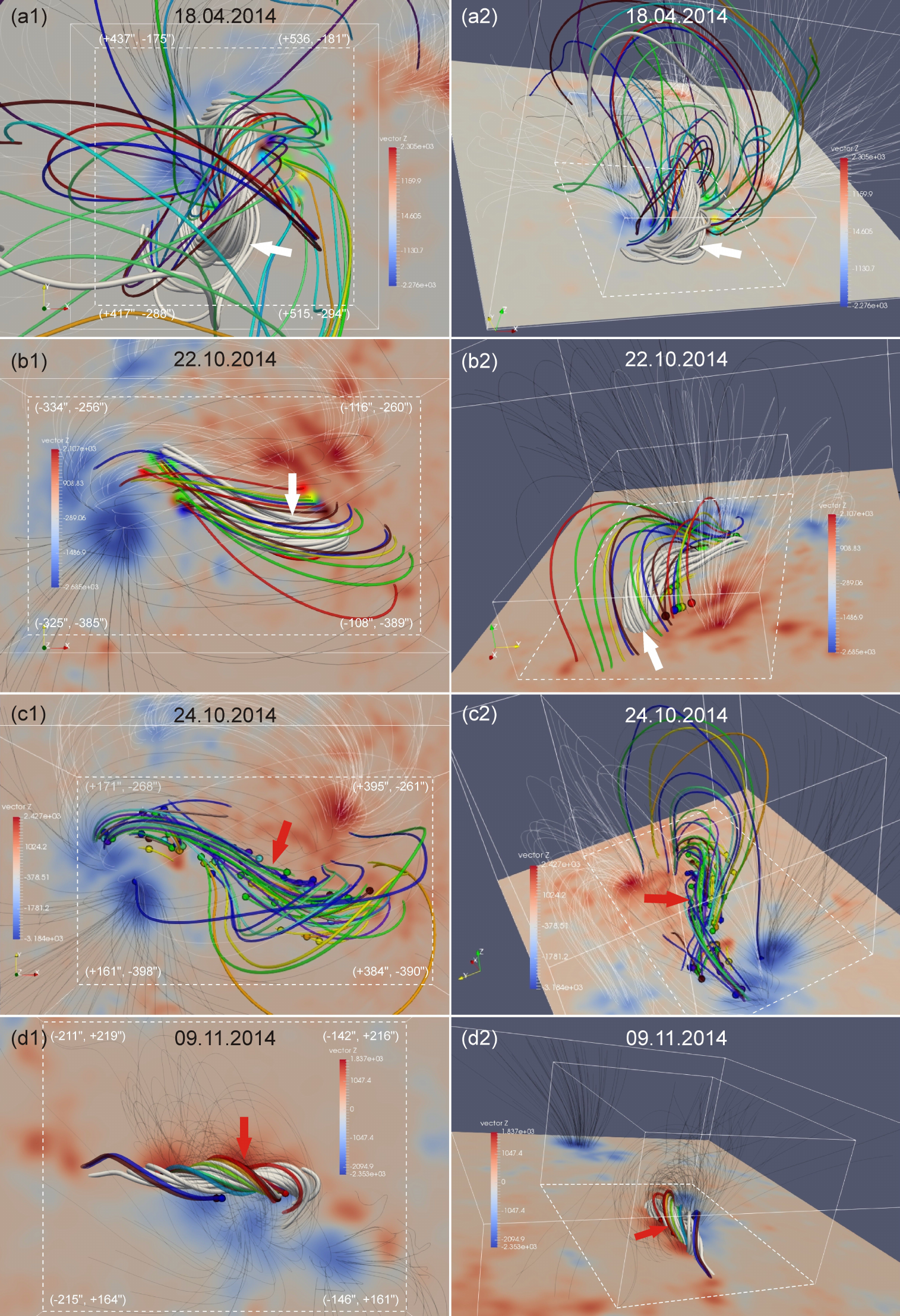}}

\caption{Reconstructed magnetic field lines and the maximum brightness centers of the sources of the HXR pulsations in the regions of the 18-Apr-2014 (a), 22-Oct-2014 (b), 24-Oct-2014 (c) and 09-Nov-2014 (d) solar flares accompanied by the HXR pulsations. The same notations as in Figure~\ref{fig:fl1}. See also Figure~\ref{fig:HXRlc} for the colors notation of the HXR peaks.} 
\label{fig:fl2}
\end{figure}

\begin{figure} 
\centerline{\includegraphics[width=0.9\textwidth]{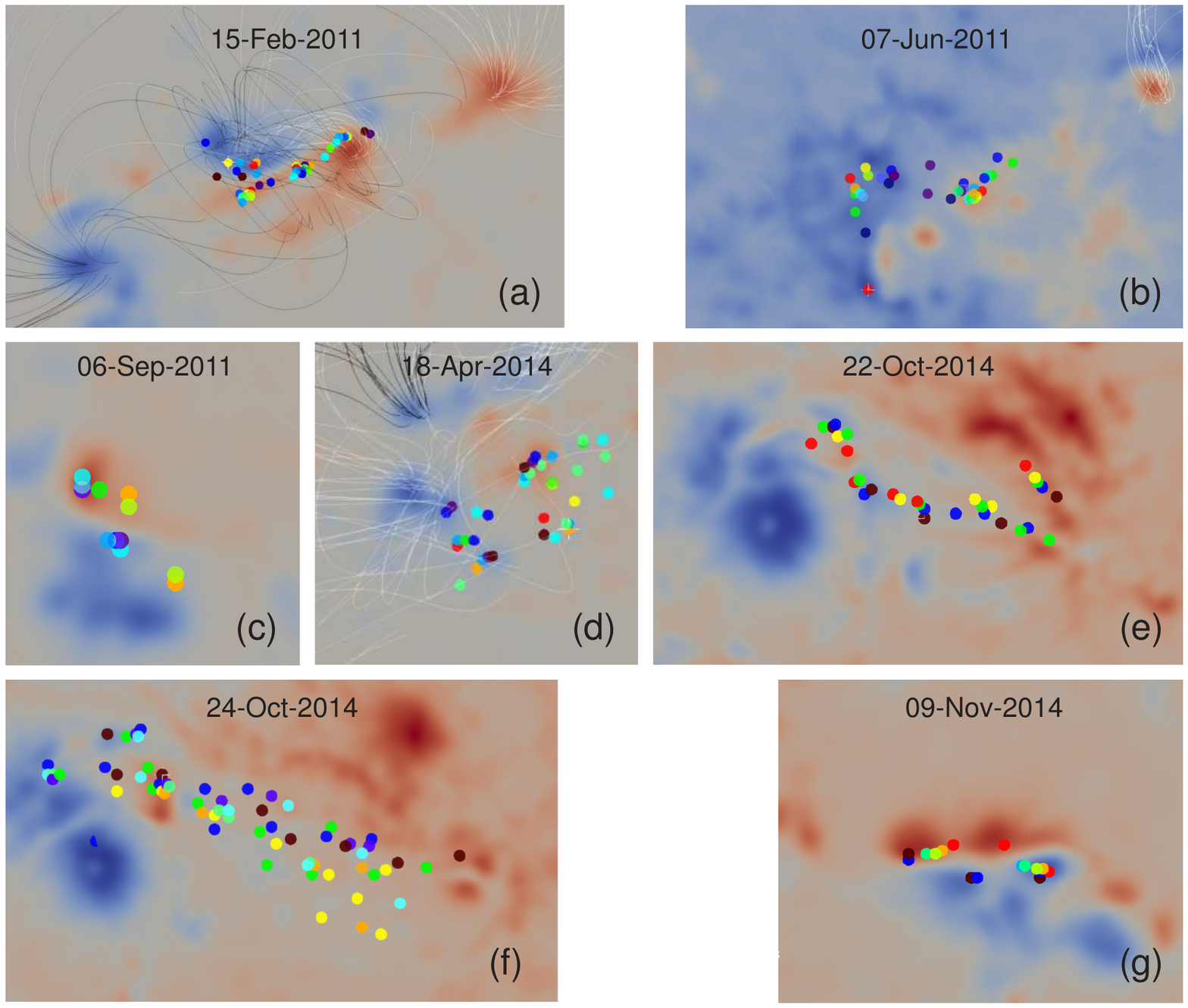}}

\caption{Positions of the maximum brightness centers of the sources of the HXR pulsations overlaid on the maps of the radial  magnetic field component ($B_{r}$) on the photosphere made with the pre-flare SDO/HMI magnetograms for the studied flare regions. The regions shown here correspond to the regions of interest shown by the thin white dashed quadrangles in Figures~\ref{fig:fl1} and~\ref{fig:fl2}. The centers of the HXR sources and their colors correspond to those ones shown in Figures~\ref{fig:fl1} and~\ref{fig:fl2} (see also Figure~\ref{fig:HXRlc} for the colors notation of the corresponding HXR peaks). } 
\label{fig:HXRonly}
\end{figure}

To visualize the extrapolated magnetic field and to compare its structure (geometry) with the location of the sources of the HXR pulsations in each event we implemented the following procedure:
\begin{enumerate}
 \item{First, we converted the HPC coordinates of the HXR sources to the Stonyhurst heliographic (HG) coordinates using the SolarSoft WCS routines (\citealp{2006A&A...449..791T}).}
 \item{Second, using the formula for the solar differential rotation derived by \cite{1990SoPh..130..295H}, we rotated the HXR sources to the times of the used pre-flare magnetograms to compensate for the time differences between the different data sets. The HXR sources were reconstructed with the use of the RHESSI data and the CLEAN image synthesis algorithm (\citealp{2002SoPh..210...61H}) for almost the same time intervals, \ie for the same HXR ($25-50$~keV) pulsations, as in Paper~I. For most of the time intervals (corresponding to the HXR pulsations) we used data of all nine RHESSI detectors and the image pixel size is $1^{\prime \prime}$.} 
\item{Third, suggesting that the synthesized HXR sources ($25-50$~keV) of the HXR pulsations are located, as usual (\citealp{2002SoPh..210..383A}), in the chromospheric footpoints of the flare magnetic flux tubes (field lines), we found the pixels of the magnetograms (the coordinates of which were also pre-transformed into the HG system) corresponding to the brightest pixels of the HXR sources (\ie its brightness maxima). These pixels are used as the starting points for the reconstruction of magnetic field lines in the flaring regions. They are shown by the small colored circles in Figures~\ref{fig:fl1}--\ref{fig:HXRonly}. Different colors of the circles (and the field lines started from them) correspond to different time intervals, \ie different HXR pulsations (see Figure~\ref{fig:HXRlc}; almost the same colors were used in Paper~I). The field lines are started from the heights of $H_{HXR}\approx 0.7-2.2$~Mm above the photosphere to satisfy our aforementioned assumption that the studied HXR sources are located in the chromosphere.} 
 \item{The radial component of the magnetic field on the photosphere, the reconstructed magnetic field lines and the corresponding HXR sources of the HXR pulsations (their centers of maximum brightness) are represented in Figures~\ref{fig:fl1}--\ref{fig:fl2} built using the ParaView application\footnote{\url{https://www.paraview.org/}}. The positions of the centers of the HXR sources are also shown in Figure~\ref{fig:HXRonly} without the reconstructed field lines for better clarity. Since the HXR sources are located low in the chromosphere at a height of only $H_{HXR} < 2.2$ Mm, as indicated above, the projection effect ($\Delta_{max} < H_{HXR} \times \tan{50^{\circ}} \approx 2.6$ Mm) of superimposing the HXR sources on the photospheric magnetic maps can be neglected.} 
\end{enumerate}

In addition to the magnetic field lines started from the HXR sources, we also constructed two other sets of field lines: 
\begin{enumerate}
 \item[(1)]{For all ARs studied, except the SOL2011-02-15 and SOL2014-10-24 events (see below), we tried to find localized elongated bundles of helical field lines twisted (more than once) around the common central axis. Such bundles of field lines can be considered as an approximation of an MFR (\citealp{2006SSRv..124..131G,2015JApA...36..157F,2017ScChE..60.1383C,2017ScChE..60.1408G}). Looking ahead, we note here that we found such bundles of field lines in the central part of all seven considered flare regions. They are shown by the thick light gray field lines (tubes) and also marked by the thick arrows (white or red) in Figures~\ref{fig:fl1}--\ref{fig:fl2}. We did not try specially to find (and visualize) MFRs for the SOL2011-02-15 and SOL2014-10-24 events, since the reconstructed field lines started from the HXR sources represent such elongated twisted magnetic structure by themselves (see Figure~\ref{fig:fl1}(a) and Figure~\ref{fig:fl2}(c), where an MFR is indicated also by the thick red arrow), \ie they are different components (threads) of an MFR.} 
 \item[(2)]{We also reconstructed multiple magnetic field lines started from the main (strongest) magnetic sources (the sunspots' umbras and penumbras) on the photosphere in the ARs studied. This is just to identify better the general magnetic structure of the ARs. These field lines are shown by the thin black and white curves in Figures~\ref{fig:fl1}--\ref{fig:fl2}.}
\end{enumerate} 
%%%%%%%%%%%%%%%%%%%%%%%%%%%%%%%%%%%%%%%%%%%%%%%%%%%

%%%%%%%%%%%%%%%%%%%%%%%%%%%%%%%%%%%%%%%%%%%%%%%%%%%%%%%%%%%%%%
\section{Summary of the results and discussion}\label{S-sumres}	

Here we summarize and discuss the main results of the performed analysis of seven flares studied. We remind that these flares were taken from Paper~I (\citealp{2016SoPh..291.3385K}) for the follow-up investigation. They were accompanied by sequences of HXR bursts (pulsations). These flares were selected from Paper~I because they all occurred in the ``SDO's epoch'', \ie the photospheric vector magnetograms are available for their parent ARs for the times just before their onset. This allowed us to make extrapolation of magnetic field in these ARs in the NLFFF approximation, and to investigate its structure in relation to the sources of the HXR pulsations.

Based on the performed extrapolation, first, we found that there is a spatially localized bundle of magnetic field lines twisted around their common axis (some of them --- slightly more than once), and elongated mainly along an MPIL in the core of the parent AR before each flare studied (see Figures~\ref{fig:fl1}--\ref{fig:fl2}). Such bundles of the intertwined field lines can be considered as a reliable signature of the presence of an MFR (\eg, \citealp{2009AdSpR..43..739S,2013AdSpR..51.1967S,2016ApJ...818..148L,2017ScChE..60.1383C,2017ScChE..60.1408G}, and references therein) in these ARs. Second, it can be seen clearly in Figures~\ref{fig:fl1}--\ref{fig:fl2} that the sources of different HXR pulsations are located in footpoints of different magnetic field lines. Definitely, different HXR pulsations are emitted from different flux tubes (loops) rather than from a single flare loop in those events. Both these findings are consistent with conclusions made in Paper~I (see also Introduction). 

%%%%%%%%%%%%%%%%%%%%%%%%%%%%%%%%%%%%%%%%%%%%%%%%%%%%%%%%%%%%%%
\subsection{Two cases of different magnetic geometry}\label{SS:cases}	

It is interesting to note that in two out of seven events (namely, SOL2011-02-15 and SOL2014-10-24) the reconstructed magnetic field lines initiated from the sources of the HXR pulsations form a twisted MFR (shown by the thick red arrow in Figures~\ref{fig:fl1}(a),~\ref{fig:fl2}(c)) by themselves. These field lines are a part of an MFR~--- they are some of its individual fibers (or threads). We will call this as case A.  

However, in the remaining five events (SOL2011-06-07, SOL2011-09-06, SOL2014-04-18, SOL2014-10-22 and SOL2014-11-09) the field lines reconstructed from the HXR sources are not directly a part of an MFR situated in a parent AR (shown by the thick arrow in Figures~\ref{fig:fl1}(b,c),~\ref{fig:fl2}(a,b,d)). These field lines are located around an MFR in the form of an overlying arcade of magnetic loops. We will call this situation as case B.

Here it should be noted that the conclusion about whether the field lines are part of an MFR or not is rather subjective. This is illustrated by the SOL2014-11-09 event (Figure~\ref{fig:fl2}(d)), in which the parent field lines of the HXR pulsations entangle the MFR quite tightly, without a significant gap, as, say, in the SOL2014-04-18 (Figure~\ref{fig:fl2}(a)) and SOL2014-10-22 (Figure~\ref{fig:fl2}(b)) events. This may indicate that there is no fundamental difference between the events of cases A and B, and the energy release processes may be similar in both cases.
%%%%%%%%%%%%%%%%%%%%%%%%%%%%%%%%%%%%%%%%%%%%%%%%%%%%%%%%%%%%%%

%%%%%%%%%%%%%%%%%%%%%%%%%%%%%%%%%%%%%%%%%%%%%%%%%%%%%%%%%%%%%%
\subsection{Interpretation of the eruptive events}\label{SS:erupt}	

Four of the studied events (SOL2011-02-15, SOL2011-06-07, SOL2011-09-06, and SOL2014-04-18) were definitely accompanied by eruptions and CMEs, \ie they can be classified as the eruptive events. Studies of various aspects of these eruptive events can be found in the multiple papers (\eg, \cite{2011ApJ...738..167S,2014ApJ...788...60J} for SOL2011-02-15; \cite{2012ApJ...745L...5C,2013ApJ...777...30I,2017Ge&Ae..57.1067K} for SOL2011-06-07; \cite{2014ApJ...780...55J,2016A&A...591A.141J} for SOL2011-09-06; \cite{2015ApJ...804...82C,2015ApJ...810...45B,2016ApJ...833...87C} for SOL2014-04-18). 

The ``standard'' 3D solar flare model is applicable for their interpretation (\citealp{2011LRSP....8....6S,2015SoPh..290.3425J}). In this case, the HXR pulsations can be a result of successive episodes of magnetic reconnection, acceleration of electrons and their precipitation along different magnetic flux tubes (loops) of a magnetic arcade non-uniformly (along its longitudinal axis or an MPIL) stretched by an erupting MFR (see, \eg, discussions in \citealp{2005ApJ...625L.143G,2010ApJ...721L.193L}). In particular, the ``zipping-like'' or ``whipping-like'' asymmetric MFR eruption \citep{2009ApJ...691.1079L} could explain the apparent motion of the HXR sources along the MPIL observed in the impulsive phase of the flares studied (see Table~4 in Paper~I). In this discussion we do not address the question of what determines the speed of the HXR source progression along the solar surface in this scenario, which should be a subject of a dedicated theoretical modeling. The absence of visible HXR sources in the footpoints of some overlying field lines, surrounding an erupting MFR, may be due to the lower efficiency of magnetic reconnection and electron acceleration in these flux tubes, and insufficient sensitivity and dynamic range of RHESSI.

On the other hand, we cannot totally rule out other possibilities. In particular, it is not excluded that successive episodes of magnetic reconnection and particle acceleration (consequently, HXR pulsations) may occur in the interaction regions of an MFR's outer shells with different surrounding magnetic loops. From Figure~\ref{fig:fl1}(b,c) and Figure~\ref{fig:fl2}(a) it can be seen that the orientation of the MFR's field lines (thick light gray) is different from the orientation of the surrounding field lines hosting the sources of the HXR pulsations (color). Such different orientation of the interacting magnetic flux tubes is a favorable condition for the initiation of magnetic reconnection (\eg, \citealp{1996SSRv...77....1S,2001ApJ...553..905L}). Further observations and modeling are required to confirm or disprove this scenario (see also \citealp{2012A&A...548A..89N,2016ApJ...832..106H}). 

It should be noted that eruption of MFRs is a dynamical phenomenon, and the magnetic structure of ARs changes during eruption and associated flares because of magnetic reconnection. However, we believe that the extrapolated pre-flare magnetic field lines approximate the general magnetic structure of the ARs studied quite realistically and reliably. The pre-eruptive (pre-flare) magnetic structure determines the general dynamics of the following eruption, the energy release processes and the flare emission sources (\eg, \citealp{2017ScChE..60.1408G,2018Natur.554..211A}). This has been demonstrated in numerous previous works, including interpretation of the appearance and location of flare ribbons and HXR sources (see, \eg, \citealp{1988SvA....32..308G,1994SoPh..150..221D,2004ARep...48..246S,2010ApJ...721L.193L,2014ARep...58..488D,2014ApJ...788...60J,2016ApJ...818..168I,2016A&A...591A.141J}). Apparently, the spatial dynamics of the HXR sources in eruptive events is determined by the development of an erupting MFR. However, the magnetostatic approach can still be used for our study for the following reason. As we see from the observations, episodes of reconnection happened successively in different loops at different times. This means that at each specific time (\ie, during specific HXR peak), the magnetic field changed its topology only locally, in the vicinity of some specific field lines, while no other drastic changes occurred in other field lines of the parent AR at that time. Our analysis only roughly shows at which field lines the energy release could occur during the flares. In this study, we do not attempt to explain in details the observed dynamical behavior of the HXR sources, \ie to explain the exact reason of appearance of the HXR sources in the specific places in specific times, and the ordering of this appearance. We only demonstrate that the sources of different HXR pulsations appeared at footpoints of different magnetic field lines, which are not a part of the same magnetic loop.

The discussed concept of generation of HXR (and other wavebands) pulsations does not, in general, require the presence of MHD waves and oscillations, as it is often assumed (see Introduction). In this concept, pulsations are just a result of a triggering of energy release and acceleration of particles in certain different magnetic elements (flux tubes) that are somehow different from the neighboring magnetic flux tubes, due to non-uniform, essentially 3D, evolution of an MFR in highly inhomogeneous medium of parent ARs, consisting of multitude of magnetoplasma elements with different physical parameters. Different mechanisms can accelerate particles during such eruption of an MFR (\citealp{2002SSRv..101....1A,2011SSRv..159..357Z}). One of the most promising mechanisms is related to the multiple coalescence of magnetic islands formed beneath an erupting MFR as a result of the fast magnetic reconnection (\citealp{2006Natur.443..553D,2016ApJ...820...60G}). Numerical simulations show that such process can be ``bursty'' and ``patchy'' (\eg, \citealp{2000A&A...360..715K,2006GeoRL..3313105D,2011ApJ...737...24B}). This may explain why flare pulsations sometimes have a rather random character than show quasi-periodic behavior (\eg, \cite{2011A&A...533A..61G,2015ApJ...798..108I,2016ApJ...833..284I}, and Paper~I). 
%%%%%%%%%%%%%%%%%%%%%%%%%%%%%%%%%%%%%%%%%%%%%%%%%%%%%%%%%%%%%%

%%%%%%%%%%%%%%%%%%%%%%%%%%%%%%%%%%%%%%%%%%%%%%%%%%%%%%%%%%%%%%
\subsection{Interpretation of the non-eruptive or confined events}\label{SS:conf}	

Three of the studied events (SOL2014-10-22, SOL2014-10-24, and SOL2014-11-09) were not accompanied by perceptible eruptions and CMEs (see, \eg, \cite{2015ApJ...808L..24C,2015ApJ...804L..28S,2015ApJ...801L..23T,2016ApJ...826..119L} for SOL2014-10-22 and SOL2014-10-24; Paper~I and \cite{2016ApJ...823L..13L} for SOL2014-11-09). Thus, these flares can be classified as the non-eruptive events. Since these events were located near the center of the solar disk, the possibility remains that the slight (non-detectable) rise of the MFR could still occur at the initial stage of the development of these events, but due to some reasons did not develop into a full eruption and CME (see \cite{2016ApJ...818..168I,2018Natur.554..211A} for the modeling of the SOL2014-10-24 event and discussions in the papers cited above). In this case, these events can be considered as the confined flares (such term is used, \eg, by \cite{2015ApJ...808L..24C,2015ApJ...801L..23T}).

If a slight rise of the MFR indeed occurred in these events (in particular, the onset of the kink instability of the MFR in the SOL2014-10-24 event is shown by \cite{2018Natur.554..211A}), it could be accompanied by formation of a current sheet beneath the rising MFR (\eg, \citealp{1990JGR....9511919F,2000ApJ...545..524C}). In such case, the interpretation of the non-uniform energy release processes along the MPIL can be almost the same as in the case of the ``standard'' 3D eruptive flares discussed in the previous sub-section. For instance, the ``zipping-like'' asymmetric rise of the MFR is possible, which could result in the propagation of the reconnection front along the current sheet (\ie, along the MPIL) and successive episodes of energy release in different flux tubes. Similar effect is also expected in frames of the models considering some types of instabilities of a current sheet extended along the MPIL (\citealp{2012SoPh..277..283A,2016AstL...42..841L}). Another possibility is to trigger multiple episodes of magnetic reconnection in different parts of a current sheet by the propagating slow magneto-acoustic waves as discussed by \cite{2011ApJ...730L..27N}. It is worth mentioning that a current sheet can be also created in the outer shells of a rising MFR during its interaction with the overlying magnetic arcade (\eg, \citealp{2016ApJ...832..106H}). The aforementioned processes of initiation of the non-uniform reconnection could be also happen in such helical current sheet, and this may explain some of the case B events.

However, it is not excluded that there was no rise of the MFR at all in these events. To interpret the observations in such a situation, it is possible to use the model of the 3D ``zipper reconnection'' developed recently by \cite{2017SoPh..292...25P}. This model is designed for two different magnetic configurations: (1) with and (2) without a pre-existing MFR under a sheared magnetic arcade. The pre-existence of an MFR is not necessary, it can be formed during a flare due to magnetic reconnection. Since we found the MFRs in the flare regions studied, they rather correspond to the model magnetic configuration (1). The eruption of an MFR is not necessary (although it is possible in this model). It can be held by, \eg, the magnetic tension of the overlying loops (\citealp{2014masu.book.....P,2015Natur.528..526M}). The reconnection in this model occurs non-simultaneously along the entire arcade, but successively between some pairs of nearby loops, in the sequence of multiple individual reconnection episodes, called ``simple zippets''. Starting at one pair of loops, e.g. due to initial resistive instability, this process can spread along the MPIL. If the pre-flare magnetic field is not uniformly distributed along the MPIL (what we actually observe), the reconnection process will be discrete and the observed flare radiative emission will also be inhomogeneous in time and space. This is manifested, in particular, by the sequential appearance of the HXR sources in different places along the MPIL and by a sequence of the observed HXR pulsations. 

The energy release processes and formation of the X-shaped flare ribbons in the SOL2014-11-09 confined flare was discussed by \cite{2016ApJ...823L..13L} based on the magnetic field reconstruction and the concept of the 3D reconnection at a separator. However, the HXR pulsations were not discussed in that paper. The ``zipping reconnection'' concept could be incorporated in the proposed model to interpret the dynamics of the HXR sources. 
%%%%%%%%%%%%%%%%%%%%%%%%%%%%%%%%%%%%%%%%%%%%%%%%%%%%%%%%%%%%%% 

%%%%%%%%%%%%%%%%%%%%%%%%%%%%%%%%%%%%%%%%%%%%%%%%%%%%%%%%%%%%%%
\subsection{On the lack of connectivity of the HXR sources}\label{SS:connect}	

One can see from Figures~\ref{fig:fl1},~\ref{fig:fl2} that there is an apparent lack of connectivity of the HXR sources of the same colors, \ie the footpoints of the field lines are mainly associated with only one HXR source, and not to the conjugated source (of the same color) appeared at a given time. This may indicate that the NLFFF extrapolations are not entirely reconstructing the real magnetic field or that one of the HXR sources at a given time is too weak compared to the other. The RHESSI’s dynamic range is about 10, and so it may have missed the conjugated sources.

There is another possibility, how one can explain the lack of connectivity of the HXR sources appeared on opposite sides of the MPIL at a given time. The magnetic extrapolations were done based on the pre-flare magnetograms. During a flare, the connectivity of magnetic field lines could (and should) change because of the magnetic reconnection (\eg, \citealp{2015NatCo...6E7598S,2016ApJ...833..221W,2016NatCo...711837X,2017ApJ...847..124H}). In this situation, paired HXR sources appeared simultaneously could not be connected by a field line reconstructed from a pre-flare magnetogram. Moreover, HXR sources on different sides of an MPIL do not have to be connected by the same field line at all, since electrons generating these HXR sources may be accelerated and injected into different field lines, even in the case of a common initial region of energy release and acceleration. Any real acceleration region has a finite size, while a field line is an abstraction having a zero radius of cross-section.
%%%%%%%%%%%%%%%%%%%%%%%%%%%%%%%%%%%%%%%%%%%%%%%%%%%%%%%%%%%%%%

%%%%%%%%%%%%%%%%%%%%%%%%%%%%%%%%%%%%%%%%%%%%%%%%%%%%%%%%%%%%%%
\subsection{Possible interpretation of the quasi-periodicity}\label{SS:qpp}	

Possible interpretation of the quasi-periodicity of the HXR pulsations within the non-MHD-wave concept was given in Paper~I. In the simplest case, this requires the constancy of an MFR speed ($v \approx const$) and the presence of spatial inhomogeneity in the physical parameters of surrounding magnetic arcades, with a characteristic spatial scale ($l \approx const$) along the direction of the MFR motion. In such case, the quasi-period of pulsations can be determined simply as $P \approx \left\langle P \right\rangle \approx l/v \approx const$. However, this interpretation has some shortcomings too. First, it is unclear why an MFR should move at a constant speed along the MPIL. Second, the nature of the spatial modulation of the inhomogeneous surrounding magnetic structures is also not obvious. There is a possibility that a quasi-periodic modulation may exist along an MFR itself, say, due to some oscillations/waves in it~--- similar to prominence oscillations (\eg, \citealp{2002SoPh..206...45O,2009SSRv..149..175O}). This hypotheses, however, requires further study, and is beyond the scope of the present work, which will be addressed elsewhere. In addition, the spatial quasi-periodicity could, in particular, result from some instabilities of a reconnecting macroscopic current sheet (\eg, \citealp{2012SoPh..277..283A,2016AstL...42..841L}) created during an MFR rise, or from the corrugation instability of a coronal arcade recently described by \cite{2017SoPh..292..184K}. 

In the non-eruptive flares, when the MFR is at rest, in order to explain the quasi-periodicity of the HXR pulsations in the frames of the 3D ``zipper reconnection'' (\citealp{2017SoPh..292...25P}), it is also necessary to suggest a spatial inhomogeneity in the magnetic arcade and a constancy of the speed of the reconnection trigger movement along the MPIL. What is a trigger and why it should have approximately constant speed (usually below the sound and Alfv\'en speeds in the corona; see, \eg, Paper~I) is not yet clear. In addition, the detected variation of the HXR emission could be somehow linked with the vertical oscillations of an emerging MFR detected by \citep{2014ApJ...797L..22K}, which would modulate the local inflow rate in the reconnection. However, theoretical modeling of this effect is still absent.

We would like to note also that the discussed concept does not reject the possibility that other physical processes, in particular, related to MHD waves and oscillations, may also act in flare regions and cause quasi-periodic component of pulsations in some cases. In any case, our findings indicate an important role of a pre-flare magnetic structure (in particular, an MFR), which is more complex than a ``single'' loop, and the three-dimensional character of the development of energy release in such magnetic structure, in the time variability of the flaring emission. This is what we originally wanted to show by this work.
%%%%%%%%%%%%%%%%%%%%%%%%%%%%%%%%%%%%%%%%%%%%%%%%%%%%%%%%%%%%%%

%%%%%%%%%%%%%%%%%%%%%%%%%%%%%%%%%%%%%%%%%%%%%%%%%%%%%%%%%%%%%%
\subsection{Geometrical characteristics of the found MFRs}\label{SS:geom}	

\begin{figure} 
\centerline{\includegraphics[width=0.99\textwidth]{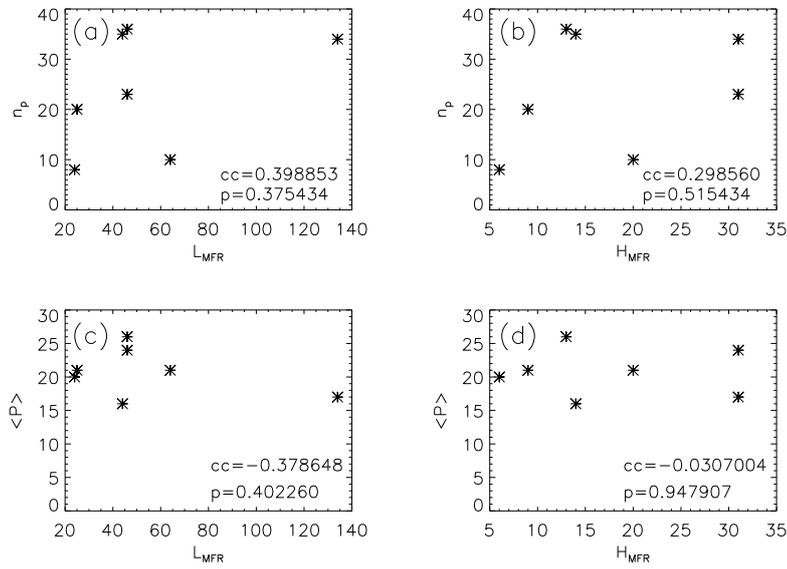}}

\caption{Scatter plots of characteristics of the HXR pulsations ($n_{p}$ and $\left\langle P \right\rangle$) and geometrical characteristics of the reconstructed magnetic flux ropes ($L_{\mathrm{MFR}}$ and $H_{\mathrm{MFR}}$). The values of the linear Pearson correlation coefficient ($cc$) and $p$-values are shown in the right bottom corners.} 
\label{fig:scatterplot}
\end{figure}

Geometrical characteristics of the found MFRs (the length of its central axis, $L_{\mathrm{MFR}}$, and height of its top above the photosphere, $H_{\mathrm{MFR}}$) are summarized in Table~\ref{tab:mfr}. These characteristics were estimated roughly by visual analysis of the reconstructed magnetic field lines using the ParaView application. By an MFR we mean a set of magnetic field lines bounded in space and winding coherently on some imaginary axis stretched along the MPIL. We realize the difficulty to determine an MFR's boundaries just by visual inspection of the magnetic field lines. However, there is no need for a more accurate evaluation of these parameters in this study. In Table~\ref{tab:mfr} we also present the number ($n_{p}$) of significant HXR ($25-50$~keV) pulsations together with the average time differences ($\left\langle P \right\rangle$) between the peaks of successive pulsations, taken from Table~2 of Paper~I. The linear Pearson correlation coefficient ($cc$) and $p$-value calculated for the pairs of the corresponding physical variables are shown in the right bottom corners of each panel on Figure~\ref{fig:scatterplot}. All calculated $p$-values are not close to zero. This indicates that correlation between the pairs of the physical variables cannot be considered as significant. Due to this reason, we refrain from discussing possible causes of weak correlations of the physical characteristics of the MFRs and HXR pulsations. Data statistics are insufficient (and have to be extended in future works) to draw any physical conclusions.

The last two columns of Table~\ref{tab:mfr} contain information about the group (see Paper~I) and case (see sub-section~\ref{SS:cases}) of each event. We recall here that the group 1 flares show systematic displacement of the HXR sources from pulsation to pulsation with respect to a MPIL, which has a simple extended trace on the photosphere, and the group 2 flares show more chaotic displacements of the HXR sources with respect to a MPIL having a more complicated structure. One can see from Table~\ref{tab:mfr} that there is no consistency between these two different classifications of the flare regions studied. This indicates that the magnetostatic approach used in this work, probably, is not enough to describe all details of the spatio-temporal evolution of the flare energy release processes. This requires more advanced modeling taking dynamics of a flare region into account.
%%%%%%%%%%%%%%%%%%%%%%%%%%%%%%%%%%%%%%%%%%%%%%%%%%%%%%%%%%%%%

%%%%%%%%%%%%%%%%%%%%%%%%%%%%%%%%%%%%%%%%%%%%%%%%%%%%%%%%%%%%%%
\section{Conclusion}\label{S-Concl} 

By means of magnetic field extrapolation in the NLFFF approximation we investigated the magnetic geometry and structure of seven solar flare regions accompanied by HXR pulsations. These flares were chosen from the catalog in Paper~I on the basis of the availability of vector magnetograms for their parent ARs, obtained with the SDO/HMI data. We found that there is an MFR elongated along an MPIL in the core of each AR studied, before each flare. In two flare regions the sources of the HXR pulsations are located at the footpoints of different magnetic field lines which are constituent parts of the MFR. In five remaining flare regions the parent field lines of the HXR pulsations are not a part of the MFR, but surround it in the form of an arcade of magnetic loops. These results support the concept discussed in Paper~I, namely that the HXR pulsations are a consequence of successive episodes of energy release in different magnetic flux tubes (threads) of a complex AR, possibly triggered by non-uniform evolution (full or confined eruption) of an MFR. This concept is consistent with the ``standard'' 3D model of solar flares and does not require the presence of MHD waves and oscillations in flare regions for interpreting the HXR (as well as other wavebands) pulsations. In the absence of eruption, some other mechanisms, such as the 3D ``zipper reconnection'' may lead to the non-uniform energy release in a sheared magnetic arcade with an underlying MFR. However, details of the time variability produced by these mechanisms that can be attributed to the ``magnetic dripping'' class of QPP models (\citealp{2010PPCF...52l4009N}), require further investigation.
%%%%%%%%%%%%%%%%%%%%%%%%%%%%%%%%%%%%%%%%%%%%%%%%%%%%%%%%%%%%%

%%%%%%%%%%%%%%%%%%%%%%%%%%%%%%%%%%%%%%%%%%%%%%%%%%%%%%%%%%%%%
\section*{Acknowledgements}
This work is based upon the activities of the international science team ``Pulsations in solar flares: matching observations and models'' supported by the International Space Science Institute -- Beijing, China. We are grateful to the RHESSI and SDO/HMI teams, whose data products were used in this study. We also thank the anonymous reviewers for a number of useful criticisms. This study was supported by the British Council via the Institutional Links Programme (Project 277352569 \lq\lq Seismology of Solar Coronal Active Regions \rq\rq). SAK was supported by the Russian Foundation for Basic Research (grants No.~15-02-08028, 16-32-50117, 16-02-00328) and by the Russian Science Foundation (grant No.~16-12-10448). VMN was partially supported by the HSE Teaching Excellence Initiatives. The work was also supported by the NSFC grant No.~41731070.  
%%%%%%%%%%%%%%%%%%%%%%%%%%%%%%%%%%%%%%%%%%%%%%%%%%%%%%%%%%%%%

%%%%%%%%%%%%%%%%%%%%%%%%%%%%%%%%%%%%%%%%%%%%%%%%%%%%%%%%%%%%%
%\section{Bibliography styles}
\section*{References}
\bibliography{zimovets_bib_file}
%%%%%%%%%%%%%%%%%%%%%%%%%%%%%%%%%%%%%%%%%%%%%%%%%%%%%%%%%%%%%

%%%%%%%%%%%%%%%%%%%%%%%%%%%%%%%%%%%%%%%%%%%%%%%%%%%%%%%%%%%%%
%%% Figure 1
\newpage
%%%%%%%%%%%%%%%%%%%%%%%%%%%%%%%%%%%%%%%%%%%%%%%%%%%%%%%%%%%%%

%%%%%%%%%%%%%%%%%%%%%%%%%%%%%%%%%%%%%%%%%%%%%%%%%%%%%%%%%%%%%
%%% Figure 2
\newpage
%%%%%%%%%%%%%%%%%%%%%%%%%%%%%%%%%%%%%%%%%%%%%%%%%%%%%%%%%%%%%

%%%%%%%%%%%%%%%%%%%%%%%%%%%%%%%%%%%%%%%%%%%%%%%%%%%%%%%%%%%%%
%%% Figure 3
\newpage
%%%%%%%%%%%%%%%%%%%%%%%%%%%%%%%%%%%%%%%%%%%%%%%%%%%%%%%%%%%%%

%%%%%%%%%%%%%%%%%%%%%%%%%%%%%%%%%%%%%%%%%%%%%%%%%%%%%%%%%%%%%
%%% Figure 4
\newpage
%%%%%%%%%%%%%%%%%%%%%%%%%%%%%%%%%%%%%%%%%%%%%%%%%%%%%%%%%%%%%

%%%%%%%%%%%%%%%%%%%%%%%%%%%%%%%%%%%%%%%%%%%%%%%%%%%%%%%%%%%%%
%%% Figure 5
\newpage
%%%%%%%%%%%%%%%%%%%%%%%%%%%%%%%%%%%%%%%%%%%%%%%%%%%%%%%%%%%%%

%%%%%%%%%%%%%%%%%%%%%%%%%%%%%%%%%%%%%%%%%%%%%%%%%%%%%%%%%%%%%
%% Table 1
\newpage
\begin{table}
\caption{The metrics of the magnetic fields reconstructed in the NLFF approximation for the investigated flares.}
\begin{tabular}{|c|c|c|c|c|c|c|}
\hline

Flare &  Flare & GOES & $L_{1}$ & $L_{2}$ & $\left(\textbf{j},\textbf{B}\right)$-angle, & $E_{nlff}/E_{pot}$ \\
id. & no. & class & & & deg & \\
\hline
SOL2011-02-15 & 23 & X2.2 & 0.86 & 0.31 & 7.20 & 1.31 \\
SOL2011-06-07 & 24 & M2.5 & 1.65 & 1.15 & 6.22 & 1.36 \\
SOL2011-09-06 & 25 & X2.1 & 1.43 & 0.86 & 9.62 & 1.22 \\
SOL2014-04-18 & 26 & M7.3 & 1.75 & 1.15 & 7.29 & 1.39 \\ 
SOL2014-10-22 & 27 & X1.6 & 0.46 & 0.25 & 7.94 & 1.17 \\
SOL2014-10-24 & 28 & X3.1 & 0.42 & 0.23 & 7.38 & 1.20 \\
SOL2014-11-09 & 29 & M2.3 & 1.09 & 0.72 & 9.56 & 1.36 \\
\hline
\end{tabular}
\label{tab:nlfff}
\end{table}
%%%%%%%%%%%%%%%%%%%%%%%%%%%%%%%%%%%%%%%%%%%%%%%%%%%%%%%%%%%%%

%%%%%%%%%%%%%%%%%%%%%%%%%%%%%%%%%%%%%%%%%%%%%%%%%%%%%%%%%%%%%
%% Table 2
\newpage
\begin{table}
\caption{Geometrical characteristics of the reconstructed MFRs ($L_{\mathrm{MFR}}$ --- length, $H_{\mathrm{MFR}}$ --- top height), characteristics of the HXR pulsations ($n_{p}$ --- number of pulsations, $\left\langle P \right\rangle$ --- average time difference between neighboring pulsations), and two different classifications of the flares studied.}
\begin{tabular}{|c|c|c|c|c|c|c|}
\hline
1     &   2      &      3     &   4    &   5 & 6 & 7  \\
\hline
Flare & $L_{\mathrm{MFR}}$,~Mm & $H_{\mathrm{MFR}}$,~Mm & $n_{p}$ & $\left\langle P \right\rangle$,~s & group & case \\
\hline
SOL2011-02-15  & 44 & 14 & 35 & 16 & 2 & A \\
SOL2011-06-07  & 46 & 13 & 36 & 26 & 1 & B \\
SOL2011-09-06  & 25 & 9  & 20 & 21 & 1 & B \\
SOL2014-04-18  & 46 & 31 & 23 & 24 & 2 & B \\ 
SOL2014-10-22  & 64 & 20 & 10 & 21 & 1 & B \\
SOL2014-10-24  & 134 & 31 & 34 & 17 & 2 & A \\
SOL2014-11-09  & 24 & 6  & 8  & 20 & 1 & A \\
\hline
\end{tabular}
\label{tab:mfr}
\end{table}
%%%%%%%%%%%%%%%%%%%%%%%%%%%%%%%%%%%%%%%%%%%%%%%%%%%%%%%%%%%%%
 
\end{document}